\documentclass[preprint,superscriptaddress,prb]{revtex4}
\pdfoutput=1
\usepackage{graphicx}
\usepackage{epstopdf}
\usepackage{bbm}
\usepackage{amsmath}
\usepackage{natbib}

\begin{document}
\title{Observation of Volkov-Pankratov states in topological HgTe heterojunctions using high-frequency compressibility}


\author{A. Inhofer}
\affiliation{Laboratoire Pierre Aigrain, Ecole Normale
Sup\'erieure, PSL Research University, CNRS, Universit\'e
Pierre et Marie Curie, Sorbonne Universit\'es, Universit\'e
Paris Diderot, Sorbonne Paris-Cit\'e, 24 rue Lhomond, 75231
Paris Cedex 05, France}
\author{S. Tchoumakov}
\affiliation{Laboratoire de Physique des Solides, CNRS UMR
8502, Univ. Paris-Sud, Univ. Paris-Saclay F-91405 Orsay Cedex,
France}
\author{B.A. Assaf}
\affiliation{D\'epartement de Physique, Ecole Normale
Sup\'erieure-PSL Research University, CNRS, 24 rue Lhomond,
75231 Paris Cedex 05, France}
\author{G. F\`eve}
\affiliation{Laboratoire Pierre Aigrain, Ecole Normale
Sup\'erieure, PSL Research University, CNRS, Universit\'e
Pierre et Marie Curie, Sorbonne Universit\'es, Universit\'e
Paris Diderot, Sorbonne Paris-Cit\'e, 24 rue Lhomond, 75231
Paris Cedex 05, France}
\author{J.M. Berroir}
\affiliation{Laboratoire Pierre Aigrain, Ecole Normale
Sup\'erieure, PSL Research University, CNRS, Universit\'e
Pierre et Marie Curie, Sorbonne Universit\'es, Universit\'e
Paris Diderot, Sorbonne Paris-Cit\'e, 24 rue Lhomond, 75231
Paris Cedex 05, France}
\author{V. Jouffrey}
\affiliation{Univ. Lyon, ENS de Lyon, Univ. Claude Bernard,
CNRS, Laboratoire de Physique, F-69342, France}
\author{D. Carpentier}
\affiliation{Univ. Lyon, ENS de Lyon, Univ. Claude Bernard,
CNRS, Laboratoire de Physique, F-69342, France}
\author{M. Goerbig}
\affiliation{Laboratoire de Physique des Solides, CNRS UMR
8502, Univ. Paris-Sud, Univ. Paris-Saclay F-91405 Orsay Cedex,
France}
\author{B. Pla\c{c}ais}
\email{bernard.placais@lpa.ens.fr} \affiliation{Laboratoire
Pierre Aigrain, Ecole Normale Sup\'erieure, PSL Research
University, CNRS, Universit\'e Pierre et Marie Curie, Sorbonne
Universit\'es, Universit\'e Paris Diderot, Sorbonne
Paris-Cit\'e, 24 rue Lhomond, 75231 Paris Cedex 05, France}
\author{\\ and  \\  K. Bendias}
\affiliation{Faculty for Physics and Astronomy and R\"ontgen
Center for Complex Material Systems, Universit\"at W\"urzburg,
Am Hubland, D-97074, W\"urzburg, Germany}
\author{D.M. Mahler}
\affiliation{Faculty for Physics and Astronomy and R\"ontgen
Center for Complex Material Systems, Universit\"at W\"urzburg,
Am Hubland, D-97074, W\"urzburg, Germany}
\author{E. Bocquillon}
\affiliation{Faculty for Physics and Astronomy and R\"ontgen
Center for Complex Material Systems, Universit\"at W\"urzburg,
Am Hubland, D-97074, W\"urzburg, Germany}
\affiliation{Laboratoire Pierre Aigrain, Ecole Normale
Sup\'erieure, PSL Research University, CNRS, Universit\'e
Pierre et Marie Curie, Sorbonne Universit\'es, Universit\'e
Paris Diderot, Sorbonne Paris-Cit\'e, 24 rue Lhomond, 75231
Paris Cedex 05, France}
\author{R. Schlereth}
\affiliation{Faculty for Physics and Astronomy and R\"ontgen
Center for Complex Material Systems, Universit\"at W\"urzburg,
Am Hubland, D-97074, W\"urzburg, Germany}
\author{C. Br\"une}
\affiliation{Faculty for Physics and Astronomy and R\"ontgen
Center for Complex Material Systems, Universit\"at W\"urzburg,
Am Hubland, D-97074, W\"urzburg, Germany}
\author{H. Buhmann}
\affiliation{Faculty for Physics and Astronomy and R\"ontgen
Center for Complex Material Systems, Universit\"at W\"urzburg,
Am Hubland, D-97074, W\"urzburg, Germany}
\author{L.W. Molenkamp}
\affiliation{Faculty for Physics and Astronomy and R\"ontgen
Center for Complex Material Systems, Universit\"at W\"urzburg,
Am Hubland, D-97074, W\"urzburg, Germany}

\begin{abstract}
It is well established that topological insulators sustain
Dirac fermion surface states as a consequence of band inversion
in the bulk. These states have a helical spin polarization and
a linear dispersion with large Fermi velocity. In this article
we report on a set of experimental observations supporting the
existence of additional massive surface states. These states
are also confined by the band inversion at a
topological-trivial semiconductor heterojunction. While first
introduced by Volkov and Pankratov (VP) before the
understanding of the topological nature of such a junction,
they  were not experimentally identified. Here we identify
their signatures on transport properties at high electric
field. By monitoring the AC admittance of HgTe topological
insulator field-effect capacitors, we access the
compressibility and conductivity of surface states in a broad
range of energies and electric fields. The Dirac states are
characterized by a compressibility minimum, a linear energy
dependence and a high mobility persisting up to energies much
larger than the transport bandgap of the bulk. At higher
energies, we observe multiple anomalous behaviors in
conductance, charge metastability and Hall resistance that
point towards the contribution of massive surface states in
transport scattering and charge transfer to the bulk. The
spectrum of these anomalies agrees with predictions of a
phenomenological model of VP states in a smooth topological
heterojunction. The model accounts for the finite interface
depth, the effect of electric fields including Dirac screening
and predicts the energy of the first VP state. The massive
surface states are a hallmark of topological heterojunctions,
whose understanding is crucial for transport studies and
applications.
\end{abstract}

\maketitle

\section{Introduction}

Topological materials have emerged
[\onlinecite{qizhang2011review}] as a new class of matter with
properties arising from peculiar crystal properties. These
include topological insulators (TIs) with band inversion due to
e.g. spin-orbit interaction
[\onlinecite{Bernevig2006science,Fu2007prl}], induced
topological superconductors [\onlinecite{qizhang2011review}],
and  topological semi-metals
[\onlinecite{Hosur1013crp,Xu2015science}]. They have attracted
much attention as hosts of exotic modes such as Majorana, Dirac
or Weyl fermions, depending on symmetry and dimensionality.
Topology also leaves macroscopic foot-prints such as the
Quantum Spin Hall Effect in 2D-TIs
[\onlinecite{Konig2007science}], quantum Hall and Dirac
screening in 3D-TIs [\onlinecite{Brune2014prx}], the fractional
Josephson effect in TI-SC junctions
[\onlinecite{Wiedemann2016ncomm,Bocquillon2016nnat,Deacon2016arxiv}],
and more recently the quantized Faraday rotation in 3D-TIs
[\onlinecite{Wu2016science,Dziom2017ncomm}].

 In spite of intensive studies some important  questions remain. One concerns the robustness of the topological states in
 practical implementations for quantum or classical topological electronics.
 Another  is the possibility of new members in the family of topological states and the related question of higher energy excitations.
 Indeed, as predicted in Ref.[\onlinecite{Karzig2013prx}], a
 limiting factor for the dynamics of Majorana braiding is the existence of surface states at higher energies.
One may wonder if similar high-energy states  accompany Dirac
states in 3D-TIs. Remarkably this question has been anticipated
theoretically by Volkov and Pankratov in
Ref.[\onlinecite{Volkov1985jetp}] who considered topological
hetero-junctions (THJ) between inverted and non inverted
Pb$_x$Sn$_{1-x}$Te semiconductors. They predicted the existence
of both the helical Dirac ground state, and of a series of
massive surface states, thereafter called Volkov-Pankratov
states (VPS). VPS theory is sketched below and detailed in
Ref.[\onlinecite{Tchoumakov2017arXiv}] with the important
addition of field effects. Some of the properties of a THJ will
certainly depend on the existence of VPS; let us mention the
anomalous screening properties in 3D TIs reported in
Ref.[\onlinecite{Brune2014prx}]. The present work aims at
unveiling VP states and highlighting their role in transport by
a high-frequency (RF) measurement of compressibility in
high-mobility strained HgTe 3D-TIs using a RF capacitor
geometry which is a building block of topological electronics.

The  RF compressibility  approach complements the ARPES
spectroscopy which has proven very efficient in identifying
Dirac states (see e.g.
Refs.[\onlinecite{Hsieh2008nature,Hsieh2009nature,Zhang2009nphys,Liu2015prb}]).
Indeed ARPES is not fully versatile as it is limited to
vacuum-TI interfaces. It therefore excludes TI capping and
gating which turn out to be essential to detect VPS as well as
for a future topological electronics. While massive surface
states were reported for oxidized Bi$_2$Se$_3$ and Bi$_2$Te$_3$
samples [\onlinecite{Bianchi2010ncomm,Bahramy2012ncomm}], they
were extrinsically induced by on-purpose surface doping. On the
other hand, the transport technique is sensitive to buried
interface states, even in the presence of capping and top
gating and is therefore well suited to map Dirac states and the
VPS in THJs. In fact a signature of VPS lies in their anomalous
electric field $\mathcal{E}$ response, coupled with the Dirac
nature of states in a THJ, such an electric field
susceptibility can be regarded as a pseudo magnetic response
with an effective magnetic field $\mathcal{E}/v_F$, where $v_F$
is the velocity of Dirac carriers. With $\mathcal{E}\sim
10^8\;\mathrm{Vm^{-1}}$ and $v_F\sim 10^6\;\mathrm{m s^{-1}}$,
one has access to very large fields $\mathcal{E}/v_F\sim
100\;\mathrm{T}$. Our technique is based on high-frequency
compressibility measurements introduced in
Ref.[\onlinecite{Pallecchi2011prb}]. The device is a
metal-insulator-TI field-effect capacitor (MITI-cap), with a
metallic plate acting as a DC and RF gate and the THJ acting as
a RF drain. The broad-band admittance measures the differential
capacitance per unit area $C(V_g)$ and the TI sheet
conductivity $\sigma(V_g)$ as function of the DC gate voltage
$V_g$. From $C(V_g)$ we extract the carrier density $n(V_g)$,
the quantum capacitance $C_Q(V_g)$, analog to the
compressibility $\chi=\partial n/\partial\mu= C_Q/e^2$, from
which we deduce the surface Fermi energy $\mu(V_g)$ that is
essential for VPS spectroscopy. The joint measurement of
$\sigma$ and $C_Q$ gives access to the diffusion constant
$\mathcal{D}(\mu)=\sigma/C_Q$ used here as a spectroscopic tool
of the surface states. Below the first VPS gap ($\mu\lesssim
E_1$), we find $\mathcal{D}\propto \mu$ and a large mobility
$\mu_e=2\mathcal{D}/\mu\simeq12\;\mathrm{m^{2}V^{-1}s^{-1}}$,
characteristic of the Dirac state. As chemical potential
crosses the VPS1 band edge ($\mu\simeq E_1$), an efficient
inter-subband scattering sets in [\onlinecite{Bastard1996}],
that is responsible for a drop of $\mathcal{D}$. The scattering
spectroscopy turns out to have a better resolution than
compressibility itself which is eventually blurred by residual
heavy-hole bulk contributions in cubic HgTe.

A fundamental concern is the distinction between VPS and
classical massive surface states such as reported in
Refs.[\onlinecite{Bianchi2010ncomm,Bahramy2012ncomm}]. Indeed
both are massive, doubly degenerate, and occur at similar
energies. The main distinction lies in the nature of
confinement: relativistic VPS are confined in an energy
gradient, whereas classical states are trapped in a potential
well involving electrostatic band bending. Experimentally they
can be distinguished by their electric field sensitivity as
discussed below (Section \ref{theory}) and in
Ref.[\onlinecite{Tchoumakov2017arXiv}]:  VPS are red-shifted
whereas trivial states are naturally blue-shifted as a positive
electric field enhances quantum confinement whereas a negative
one destroys the well altogether. This anomalous field-effect
is linked to the existence of a critical field
$\mathcal{E}_T=(\Delta_2+\Delta_1)/e\xi$ at which VPS collapse.
As discussed in Ref.[\onlinecite{Tchoumakov2017arXiv}] it
corresponds to the situation where the gap alignment between
the trivial and topological regions closes. Here $2\Delta_2$
(resp. $-2\Delta_1$) are the positive (resp. negative) band
gaps of the direct (resp. inverted) semiconductors, and $\xi$
is the penetration depth of the Dirac state. $\mathcal{E}_T$
sets both the energy scale of zero-field VPS subband series,
$E_{m}(0)=\sqrt{2 m \hbar v_F e\mathcal{E}_T}$ with $m$ a
positive integer, and the electric field dependence
 $E_{m}(\mathcal{E})=E_{m} (0)(1-\mathcal{E}^2/\mathcal{E}_T^2)^{3/4}$. In particular the
 $E_{1}(\mathcal{E})$ curve defines the phase boundary for Dirac screening.
 Taking $\xi\gtrsim\hbar v_F/\Delta_1\simeq5\;\mathrm{nm}$   and
$\Delta_1\simeq\Delta_2/3\simeq0.15\;\mathrm{eV}$
[\onlinecite{Baum2014prb}] for the CdHgTe/HgTe smooth THJs used
in this work, the predicted critical field and  VPS1 energy are
$\mathcal{E}_T\lesssim 1.2\;10^8\;\mathrm{Vm^{-1}}$ and
$E_{1}\lesssim \Delta_1 \sqrt{2
(1+\Delta_2/\Delta_1)}\simeq0.4\;\mathrm{eV}$. Five signatures
of the VPS are reported here : the observation of a Dirac-state
scattering peak in undoped samples at an energy
$E_{1}\simeq0.35\;\mathrm{eV}$ close to the predicted value,
the onset of a secondary type of carrier and Dirac screening
breakdown at the same energy, the observation of three
scattering peaks obeying the predicted VPS series
$E_{m}=\sqrt{m}E_1$, and their electric-field red-shift in
oxide capped doped samples.

The paper is organized as follows. In Section \ref{technics} we
describe the experimental principles including the fabrication
of high-mobility CdHgTe/HgTe THJs. Section \ref{experiment} is
a report of RF surface compressibility measured over a broad
range of electric field  in undoped and doped  HgTe samples.
Special attention is devoted to charge metastability effects
and their relationship to VPS. Observations are confirmed by
control experiments performed in Hall-bars with similar gate
stacks. Section \ref{theory} present a heuristic model of the
VPS and their electric field dependence, complemented by
$k\cdot P$ numerical calculations. The comparison between
theory and experiment in Section \ref{comparison} supports the
existence of VPS in CdHgTe/HgTe topological hetero-junctions.
We conclude in Section \ref{conclusion} with perspectives
offered by the VPS for a future topological electronics.

\section{Experimental principles} \label{technics}

\begin{figure}[ht]
\centerline{\includegraphics[width=14cm]{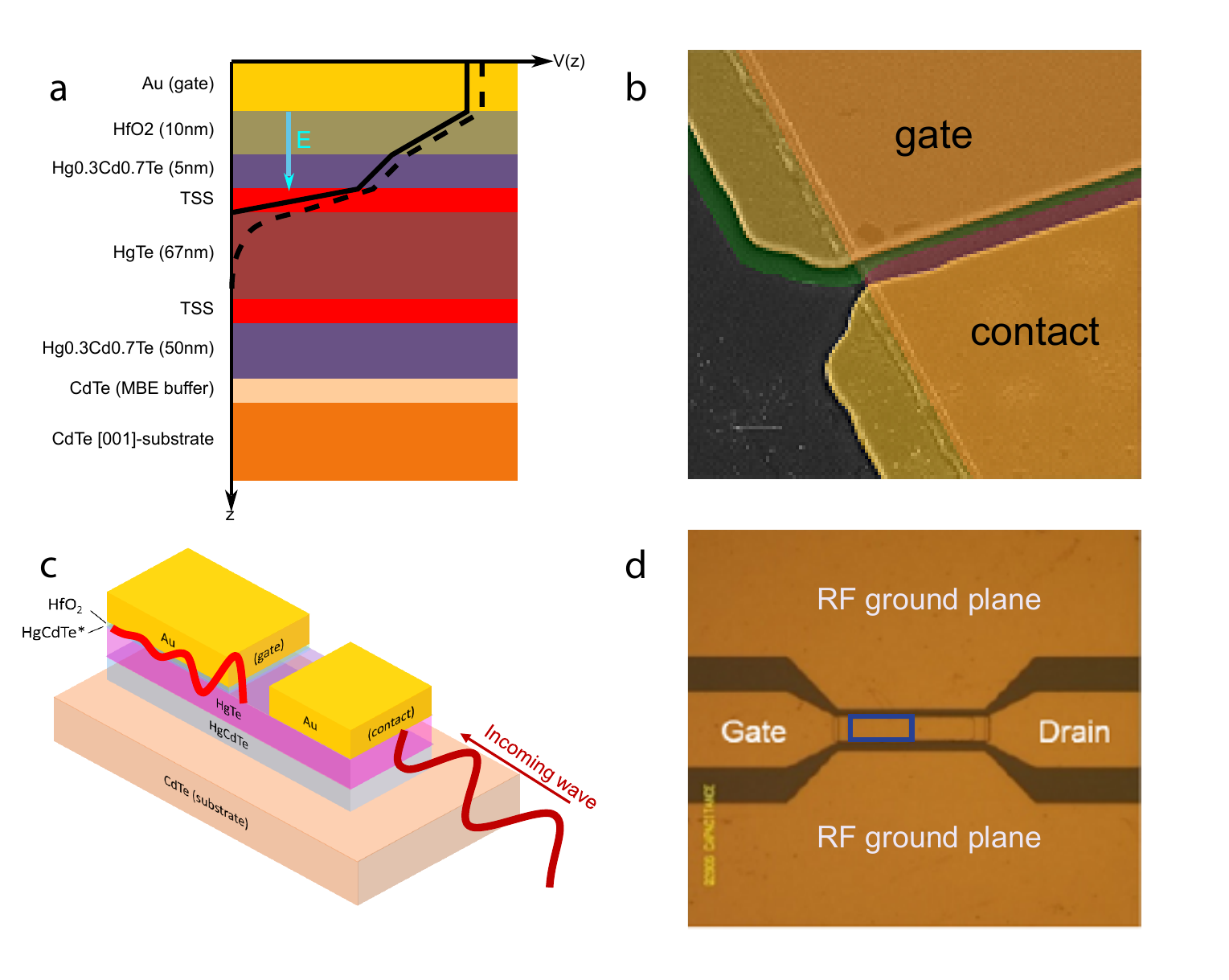}}
\caption{Description of the MITI-Caps. a) sketch of a HgTe/Cd$_{0.7}$Hg$_{0.3}$Te capped heterostructure.
Also shown in the figure are sketches of the electrostatic potential across the structure in the case of Dirac screening (solid line)
and a mixed surface/bulk state (BS) screening (dashed line). b) Colored SEM  picture at a capacitor edge showing the gate and contact
metallizations (gold), the HgTe mesa (purple) and the HfO$_2$ insulating layer (green). c) Sketch of the evanescent wave penetration
of charge in the HgTe MITI-Cap when driven at RF frequency. d) optical image of the capacitor embedded in a coplanar wave guide.
The $44\times20\;\mathrm{\mu m}$ gated area is highlighted by a blue rectangle.
}\label{fig1}
\end{figure}

The MITI-Caps are based on high mobility HgTe/CdHgTe
hetero-structures (Figure\ref{fig1}-a) grown by molecular beam
epitaxy [\onlinecite{Brune2011prl,Brune2014prx,Kozlov2014prl}]
where the HgTe layer, of thickness $t_{HgTe}=68\;\mathrm{nm}$,
is strained by the CdTe insulating substrate so as to open a
small bandgap $\simeq0.02\;\mathrm{eV}$ in the topological HgTe
layer between the light electron and heavy hole
$\Gamma_8$-bands. The gap is smaller than the light electron-
light hole ($\Gamma_8$-$\Gamma_6$) inverted bandgap,
$-2\Delta_{1}=\Delta_{\Gamma_6-\Gamma_8}\simeq-300\;\mathrm{meV}$
[\onlinecite{Guldner1973prb}], responsible for the Dirac
surface states. We use a wet etching technique
[\onlinecite{Bendias2017arXiv}] to design a mesa and optical
lithography to deposit contacts and define the gated area of
the capacitor. A gold gate electrode is evaporated on top of a
$10\;\mathrm{nm}$-thin HfO$_2$ insulating layer, grown by low
temperature atomic layer deposition (ALD) techniques. An ohmic
contact, with a resistance $R_c\simeq50\;\mathrm{Ohms}$, is
deposited by Ge/Au evaporation. A false color scanning electron
microscopy (SEM) image of the gate-contact alignment is shown
in Figure\ref{fig1}-b. Two series of samples have been
fabricated. Type $A$ samples ($S_A$) are covered by a
$5\;\mathrm{nm}$ Cd$_{0.7}$Hg$_{0.3}$Te capping layer
protecting the HgTe during the process and providing a
well-defined trivial insulator boundary of band gap
$2\Delta_2^A\simeq0.9\;\mathrm{eV}$
[\onlinecite{Baars1972ssc,Capper2010}] (Figure\ref{fig1}-a).
The total insulator thickness is $t_{ins}=15\;\mathrm{nm}$.
Type B samples ($S_{B}$)  are devoid of capping layer and
HfO$_2$ is directly deposited on HgTe (not shown in the
figure). This entails an unintentional electron doping of the
HgTe bulk (density $n_0\simeq2.6\;10^{12}\;\mathrm{cm^{-2}}$
see below). The HgTe-HfO$_2$ interface involves an oxidized
HgTe insulating layer of unknown, and presumably small, bandgap
($\Delta_2^B\ll\Delta_2^A$). We have characterized in-situ the
permittivity, $\varepsilon_\text{HfO$_2$}\simeq 3.6$, of the
thin oxide layer using a metal-oxide-metal (MOM) control
capacitor. This value is process dependent and deviates here
from to the accepted bulk value
$\varepsilon_\text{HfO$_2$}\simeq 11.7$
[\onlinecite{Neumaier2015kaprun}]. Taking
$\varepsilon_{CdHgTe}\simeq 8.5$ for the capping layer
[\onlinecite{Baars1972ssc,Capper2010}] we estimate the
capacitance of our insulating stack
$C_{ins}=2.65\;\mathrm{mF/m^2}$ in agreement with our
measurement below. The MITI-Caps are embedded in coplanar wave
guides (Fig.\ref{fig1}-d), designed for $0$--$40\;\mathrm{GHz}$
measurements. Five similar MITI-Caps of varied dimensions are
fabricated per chip complemented by MOM, dummy and thru-line
structures used for calibration purpose (see below). For
definiteness, experiments reported here refer to two capacitors
of dimensions $L\times W=44\times20\;\mathrm{\mu m}$, one of
type $A$ and one of type $B$.

Compressibility measurements have been considered so far mainly
for the characterization of Bi$_2$Se$_3$ thin crystals
[\onlinecite{Xu2015nl}], and for the purpose of Landau-level
spectroscopy in HgTe films [\onlinecite{Kozlov2016prl}]. Here
we subtract the series combination of insulator capacitance
$C_{ins}$ to the total capacitance $C_{tot}$ to obtain the
quantum capacitance correction, $C_Q=e^2\chi$.  $C_{tot}$ can
be tuned by a DC gate voltage $V_g$ controlling altogether the
charge density $n$ and the applied electric field
$\mathcal{E}$. High frequency measurements
($\lesssim10\;\mathrm{GHz}$) give access to the dissipative
regime, governed by the conductivity $\sigma$, and yield the
diffusion constant $\mathcal{D}=\sigma/C_Q$. Thin insulators
allow for
 high fields eventually exceeding the critical field $\mathcal{E}_T\simeq
1.2\;10^{8}\;\mathrm{Vm^{-1}}$ of our CdHgTe/HgTe  THJs.
Compressibility measurement plays an important role here.
First, it provides an absolute chemical potential scale with a
zero given by the Dirac dip, and secondly it gives the precise
behavior of the diffusion constant, which helps identifying the
scattering mechanisms.

 Following earlier work on graphene [\onlinecite{Pallecchi2011prb}] we primarily measure the electronic compressibility by
vectorial network analyzer (VNA) techniques (frequency range
$\omega/2\pi=50\;\mathrm{kHz}$--$8\;\mathrm{GHz}$) in a
cryogenic RF probe station at $10\;\mathrm{K}$. Standard
in-situ calibration techniques enable to de-embed the circuitry
and the contact resistance contributions from the MITI-Cap
admittance $Y(\omega)$ of interest. The admittance spectrum
$Y(\omega)$ is accurately described by a distributed
resistance-capacitance model, $Y(\omega)/W=+\jmath
C_{tot}\omega L \times\tanh(\sqrt{\jmath \sigma^{-1}
C_{tot}\omega L^2})/\sqrt{\jmath \sigma^{-1} C_{tot}\omega
L^2}$. With   $1/C_{tot}=1/C_{ins}+1/C_{Q}$, the measured
quantum capacitance per unit area $C_Q=e^2\partial n/\partial
\mu$ is a non-local compressibility where $n$ stands for the
total TI charge and $\mu$ the value of the chemical potential
at the TI surface. When two fluids are in mutual equilibrium,
e.g. Dirac and/or VPS with bulk states (BSs), they share the
same surface chemical potential and their respective densities
simply add in the total compressibility. The large thickness of
our HgTe-TI samples allows us to model them  as semi-infinite,
and the compressibility is the sum of the upper surface and a
semi-infinite bulk contribution. In the doped sample $S_B$, the
sheet conductance includes a frequency-independent BS
contribution.

\begin{figure}[ht]
\centerline{\includegraphics[width=14cm]{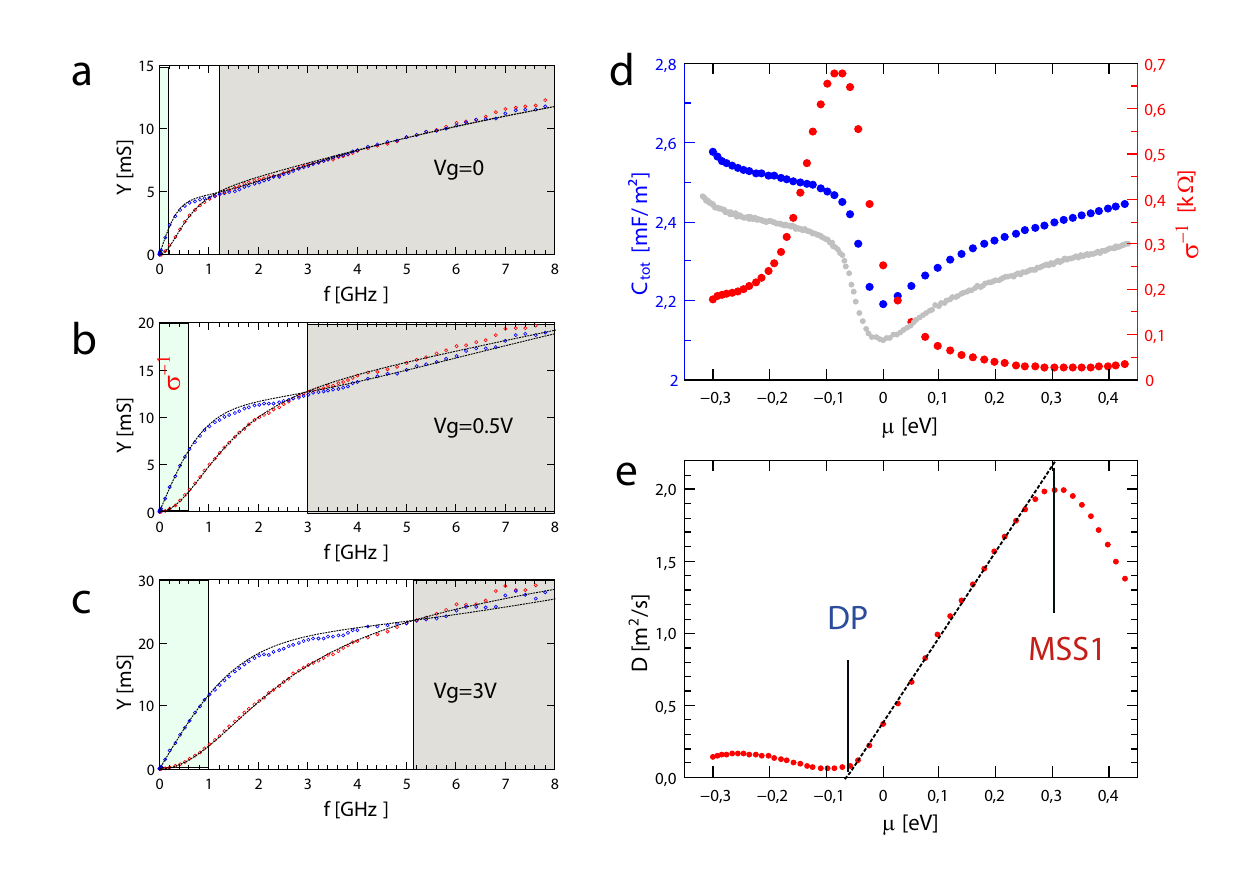}}
\caption{Scattering spectroscopy evidence of a Volkov-Pankratov state in a HgTe/CdHgTe topological heterojunction.
The quantum capacitance, conductivity and diffusion constant are deduced from capacitor RF admittance measurements.
Panel a-c) broadband spectrum of the $S_A$-capacitor complex
admittance $Y(f)$ ($\Re(Y)$ in red and $\Im(Y)$ in blue) for three typical gate voltages.
The green shaded area correspond to the quasistationary regime,
$Y/W=\jmath C_{tot}L\omega +(C_{tot}L\omega)^2\sigma^{-1} L/3$, where $C_{tot}$
and $\sigma^{-1}$ are the capacitance per unit area and the HgTe sheet resistivity. The gray shaded area is the
evanescent wave regime where $\Re(Y)\simeq\Im(Y)\simeq W \sqrt{C_{tot}\omega\sigma/2}$.
d) the quantum capacitance $C_Q$ (blue dots), deduced by de-embedding from $C_{tot}$  the insulator capacitance $C_{ins}$, and the sheet resistance $\sigma^{-1}$ (red dots),  deduced from  fits of the
AC-admittance spectra (solid lines in panel (a-c)), are plotted as function of the chemical potential as explained in the text.
The  DC measurement from Fig.\ref{fig3}-a (gray dots) is added for comparison. The resistance shows an asymmetric peak close to neutrality.
e) the diffusion constant $\mathcal{D}(\mu)=\sigma/C_Q$ shows a dip at Dirac point (DP),  a linear increase in the electron regime corresponding
to a Dirac fermion like high mobility $\mu_e=2e\mathcal{D}/\mu\simeq12\;\mathrm{m^2V^{-1}s^{-1}}$, and a peak at $\Delta\mu\approx0.35\;\mathrm{eV}$ from Dirac dip
signaling the onset of a new and efficient scattering channel. We associate the maximum
to the energy of the first Volkov-Pankratov state. Indeed the energy $\Delta\mu$ is very close to the
the theoretical prediction $E_{VP1}\approx 0.4\;\mathrm{eV}$. }\label{fig2}
\end{figure}

Typical complex admittance spectra of $S_A$  are shown in
Figs.\ref{fig2}-(a-c) for $V_g=0$ (Dirac point), $V_g=0.5$,
$3\;\mathrm{V}$ (electron doped regime). We observe three
frequency domains: a quasi-stationary domain  $Y/W=\jmath
C_{tot}L\omega +(C_{tot}L\omega)^2\sigma^{-1} L/3 $ (green
shading), an intermediate domain (unshaded)  and finally the
evanescent wave domain (grey shading), where
$\Re(Y)\simeq\Im(Y)\simeq W \sqrt{C_{tot}\omega\sigma/2}$. The
agreement between the experimental spectra and the distributed
RC line model (solid lines in Figs.\ref{fig2}-(a-c)) is
excellent and warrants the accuracy of the RF MITI-Cap approach
of compressibility. The RC line model allows extracting
reliable value of the frequency independent total capacitance
$C_{tot}$, insulator capacitance
$C_{ins}=C_{tot}(V_g\rightarrow-\infty)$, quantum capacitance
$C_{Q}(V_g)=C_{ins}C_{tot}/(C_{ins}-C_{tot})$
(Fig.\ref{fig2}-d), and conductivity $\sigma(V_g)$. Accounting
for the non-linear charge voltage characteristic we calculate
the total electron density $n =\frac{1}{e}\int
C_{tot}(V_g)dV_g$, the applied electric field
$\mathcal{E}_{ins}=n e /(C_{ins}t_{ins})$ and the surface
chemical potential, $\mu(V_g)=e\int C_{tot}/C_Q dV_g$. This
allows plotting $C_{Q}(\mu)$ and $\sigma(\mu)$ in
Fig.\ref{fig2}-d as well as the diffusion constant
$\mathcal{D}(\mu)=\sigma/C_{Q}$ used for scattering
spectroscopy in Fig.\ref{fig2}-e.

A similar chip with the same HgTe/CdHgTe hetero-structure is
produced, hosting $600\times200\;\mathrm{\mu m}$ Hall-bar
structures, with the same optical lithography wet-etching
technique equipped with a gold electrode as gate on top of
$10\,\mathrm{nm}$ thick HfO$_{2}$ insulator. Standard low
frequency AC techniques in a four point measurement
configuration are used at a temperature of $2\,\text{K}$ in
magnetic fields up to $2\,\mathrm{T}$ to measure the magnetic
field and gate voltage $V_g$  dependent longitudinal sheet
resistance $R_{xx}$ and Hall resistance $R_{xy}$. The carrier
density $n$ can be accessed in the classical Drude transport
regime by the magnetic field dependence of the Hall resistance.
Two samples of types $A$ and $B$ analog to the MITI-Cap samples
with and without capping layer (not shown) were measured.

\section{Experiment} \label{experiment}

\begin{figure}[ht]
\centerline{\includegraphics[width=14cm]{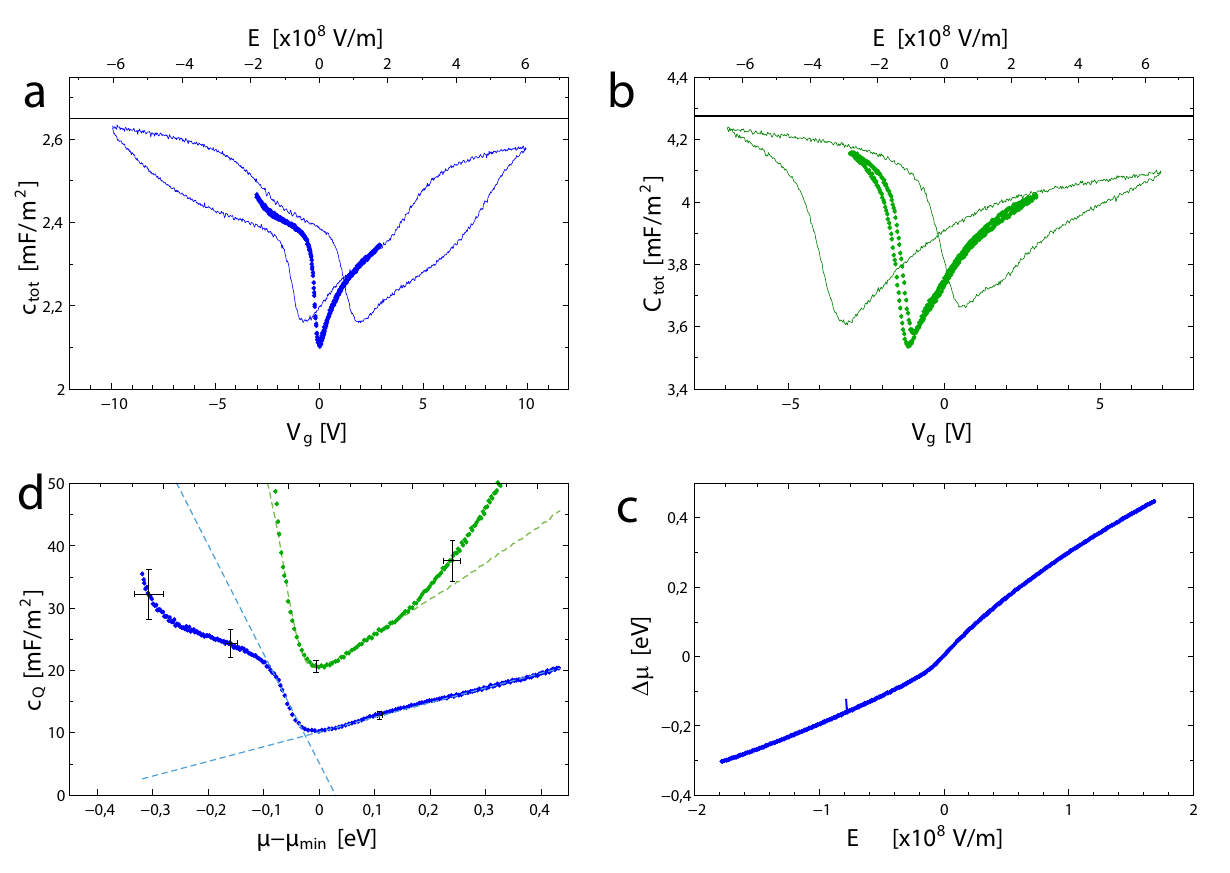}}
\caption{Reversible surface state compressibility and the breakdown of Dirac screening. Gate voltage dependence of the capacitance in the undoped
capped sample $S_A$ (panel a) and the uncapped doped sample $S_B$ (panel b) measured by lock-in techniques at $10\;\mathrm{kHz}$. In both cases we
distinguish a low electric field regime (bold lines) were $C_{tot}(V_g)$ is reversible and a high field regime
($V_g>3V$) where MITI-Cap charging is  hysteretic. Both samples show a capacitance dip in the
reversible regime corresponding to the Dirac point of the upper surface state.
It is shifted in sample $S_B$, indicating an electron-type chemical doping of density
$n\simeq 2.6\;10^{12}\;\mathrm{cm^{-2}}$. From the capacitance saturation at large gate voltage
we deduce the insulator capacitance $C_{ins}^A\simeq 2.65\;\mathrm{mF/m^2}$ and
$C_{ins}^B\simeq 4.27\;\mathrm{mF/m^2}$. Panel c) shows the surface chemical potential as
function of the applied electric field. Panel d) shows the quantum capacitance for samples
$S_A$ and $S_B$. The error bars are calculated taking a $1\%$ uncertainty in the determination
of the insulator capacitance. The main feature is the linear $C_Q(\mu)$ of sample $S_A$ in the
electron regime which is a direct signature of the intrinsic TI compressibility.
It extends over a broad  range of electrostatic doping $\sim0.3\;\mathrm{eV}$. Details of the analysis of $C_Q(\mu)$
are given in the main text.}\label{fig3}
\end{figure}

Let us first give an overview of the compressibility
measurements in samples $S_A$ and $S_B$. As shown in the
$10\;\mathrm{kHz}$ lock-in measurements of Fig.\ref{fig3}, a
striking feature is the existence of a fully reversible
compressibility in Figs.\ref{fig3}-a (and a quasi reversible
one in Figs.\ref{fig3}-b)  below a maximum gate voltage
$V_g^m=3\;\mathrm{V}$. The absence of metastability warrants
the absence of a bulk contribution consistently with the
magneto-transport measurements of
Ref.[\onlinecite{Brune2014prx}]. By contrast we observe above
$V_g^m=3\;\mathrm{V}$ a prominent charge metastability up to
the maximum amplitude, $V_g^m\gtrsim10\;\mathrm{V}$,
sustainable by the insulating dielectric stack. Metastability
signals the onset of a second type of carrier and characterizes
the breakdown of Dirac screening. The hysteresis loops have a
butterfly shape and a trend toward saturation at large negative
$V_g$. We use this asymptotic value (horizontal lines in
Figs.\ref{fig3}-a,b) to estimate the insulator capacitance
$C_{ins}$. Note that the reversible curves are eventually
history dependent themselves, the displayed ones corresponding
to the first charging cycles after cool-down to $T=10K$.
Reversible cycles subsequent to a high field sweep are
qualitatively similar but with a dip that is shifted in gate
voltage and amplitude signaling the existence of trapped bulk
carriers. We have checked that the MOM capacitance is linear
and reversible in the same conditions, confirming that
hysteresis is indeed associated to metastability in HgTe
charging.

\subsection{Compressibility of Dirac surface states}

In both samples the reversible charge response shows a
capacitance dip attributed to the Dirac surface states near
charge neutrality. In sample $S_A$ the dip is sharp and located
at $V_g=0\;\mathrm{V}$ as expected for an intrinsic TI
response. In sample $S_{B}$ the dip is shifted to
$V_g=-1.2\;\mathrm{V}$, corresponding to an n-type doping
density $n_{0}=-2.6\;10^{12}\;\mathrm{cm^{-2}}$ (bulk donor
density $N_D=4\; 10^{17}\;\mathrm{cm^{-3}}$), and smeared by
this bulk carrier contribution. On subtracting the insulator
contribution $1/C_{ins}$ and integrating the relevant
quantities, we deduce the surface chemical potential $\mu$ in
the reversible state which is plotted in Figure \ref{fig3}-c
(sample $S_A$) as a function of $\mathcal{E}_{ins}$. The
S-shape curve represents the charging path of the MITI-Cap in
the $\mu(\mathcal{E}_{ins})$ representation. Similar curves are
obtained for sample $S_B$ (not shown). Finally,
Fig.\ref{fig3}-d shows the reversible $C_Q(\mu)$ plots of $S_A$
(blue line) and $S_B$ (green line). As expected from the
additivity of compressibilities in a two-fluid system, one has
$C_ Q^B>C_Q^A$. In order to quantify the effect of bulk
carriers we have added in the figure the theoretical
expectation (green dashed line) for the additional contribution
of a trivial semiconductor to quantum capacitance with the
above estimated dopant density $N_D$. From the good
experimental agreement we conclude that the reversible
$C_Q(\mu)$ plot of sample $S_A$ is a close estimate of the
intrinsic TI response. The linear energy dependence precludes
an interpretation of compressibility in terms of conventional
massive surface states. In fact, its energy dependence can be
mapped to a Dirac fermion density of states,
$C_Q=e^2\mu/(2\pi\hbar^2v_F^2)$,  with
$v_F=1.6\;10^6\;\mathrm{ms^{-1}}$ in the electron regime and
$v_F=0.5\;10^6\;\mathrm{ms^{-1}}$ in the hole regime.  These
values are larger than the accepted
$v_F\simeq1\pm0.2\;10^6\;\mathrm{ms^{-1}}$
[\onlinecite{Liu2015prb}]. Remarkably this Dirac-like response
extends over a broad energy range ($\mu=-0.05\rightarrow
+0.30\;\mathrm{eV}$) widely exceeding the bulk transport
bandgap, consistent with the Dirac screening  reported in
Ref.[\onlinecite{Brune2014prx}]. In the data reduction we have
included a constant background $\sim 10\;\mathrm{mF m^{-2}}$
whose origin is not fully clarified. A possible explanation
would be the nesting of the Dirac point in the heavy-hole
$\Gamma_8$ branch, meaning that the observed compressibility
minimum results from a superposition of a Dirac contribution
with that of a heavy-hole $\Gamma_8$ band. \footnote{The
resurgence of a bulk contribution in the vicinity of a Dirac
point is expected however, due to the vanishing Dirac screening
at neutrality, illustrated by the divergence of the effective
screening length $\lambda=\varepsilon/C_Q$. Such effects have
been recently demonstrated in gated graphene-on-metal contacts
[\onlinecite{Wilmart2016scirep}]. The analog of TIs would be
the case of graphene-on-semiconductor, where the finite
compressibility of the small gap semiconductor is taken into
account. This case, which has not been investigated yet, is
beyond the scope of our paper that focusses on the high field
regime. In the following we shall therefore focus on the
electron-doped regime and subtract a background compressibility
$\sim0.01\;\mathrm{F m^{-2}}$.}

\subsection{Scattering spectroscopy evidence of the first VPS}

We focus here on the sample $S_A$ which is the closest
realization of an intrinsic CdHgTe/HgTe smooth THJ.
Fig.\ref{fig2}-d shows $C_{tot}(\mu)$ (blue dots) and
$\sigma^{-1}(\mu)$ (red dots) deduced from admittance spectra
in the reversible regime. The DC capacitance (grey dots) has
been reproduced for comparison. The good agreement, beside a
small amplitude shift due to calibration, shows that
compressibility is frequency-independent, confirming our
analysis of admittance spectra that dissipative effects mainly
stem from finite conductivity and not from an intrinsic
$\chi(f)$ dependence. The resistivity  exhibits a Dirac-like
peak (DP) which is slightly shifted with respect to the
capacitance dip, supporting our conjecture that the DP is
nested in the $\Gamma_8$ band. The shape of the resistance
reflects the electron-hole asymmetry of HgTe. The diffusion
constant $\mathcal{D}(\mu)=\sigma/C_Q$ in Fig.\ref{fig2}-e
shows a minimum at $\mu\simeq -0.05\;\mathrm{eV}$ and a linear
increase in the electron regime up to $\mu\simeq
0.3\;\mathrm{eV}$, corresponding to a constant mobility
$\mu_{e}=2e\mathcal{D}/\mu\simeq12\;\mathrm{m^2V^{-1}s^{-1}}$.
This large value is characteristic of massless Dirac fermions
and is comparable with values deduced from magneto-transport
experiments [\onlinecite{Brune2014prx}]. The  diffusion
constant $\mathcal{D}(\mu)$ is a marker of the scattering
mechanism, its linear dependence is indicative of a screened
charge disorder [\onlinecite{Nomura2006prl,Culcer2010prb}]. The
most remarkable feature is a drop of $\mathcal{D}(\mu)$ at
$\mu\simeq0.3\;\mathrm{eV}$. We attribute this \emph{scattering
peak} to the onset of a new scattering channel for Dirac
fermions when the surface Fermi energy
$E_1=\mu-E_{DP}\simeq0.35\;\mathrm{eV}$ crosses the bottom of a
massive subband [\onlinecite{Bastard1996}].

Sample $S_A$ being intrinsic we can rule out the possibility
that the scattering peak corresponds to a doping induced
massive surface state as reported in
[\onlinecite{Bianchi2010ncomm,Bahramy2012ncomm}]. The
scattering peak energy $E_1\simeq0.35\;\mathrm{eV}$ is very
close to the theoretical estimate for the VPS $\Delta_1 \sqrt{2
(1+\Delta_2/\Delta_1)}\simeq0.4\;\mathrm{eV}$. Identifying this
energy with that of the first VPS band edge, $E_{1}
=\sqrt{2\hbar v_F (\Delta_1+\Delta_2)/\xi}$ (see Eq.
(\ref{eq:VP}) in the theory section \ref{theory}), gives
$\xi\simeq6\;\mathrm{nm}$ (with $v_F=1.10^6\;\mathrm{m/s}$)
close to the estimate $\xi\simeq5\;\mathrm{nm}$ deduced from
numerical studies in Ref.[\onlinecite{Baum2014prb}]. From the
agreement between theoretical and experimental determinations
of the peak energy  we conclude that the scattering peak
observed in Fig.\ref{fig2}-d is a signature of the topological
VPS subband
[\onlinecite{Volkov1985jetp},\onlinecite{Tchoumakov2017arXiv}]
described in Sec. \ref{theory}. A theoretical modeling of the
peak shape, which would involve a detailed analysis of
scattering and screening mechanisms in HgTe, remains beyond the
scope of our work. $C_Q$ being monotonic in the VPS$_1$ energy
range, the peak in $\mathcal{D}(\mu)$ translates into a peak in
$\sigma(\mu)$. Fig.\ref{fig4}-a is an extension of
Fig.\ref{fig2}-d at the maximum gate voltage span, taking the
upward voltage sweep of the hysteresis loops. It shows that the
conductance peak evolves into a resistance plateau at large
electric field, and that no additional conductance peak is
observed in this sample.

\subsection{Additional signatures of VPS carriers}

\begin{figure}[ht]
\centerline{\includegraphics[width=16cm]{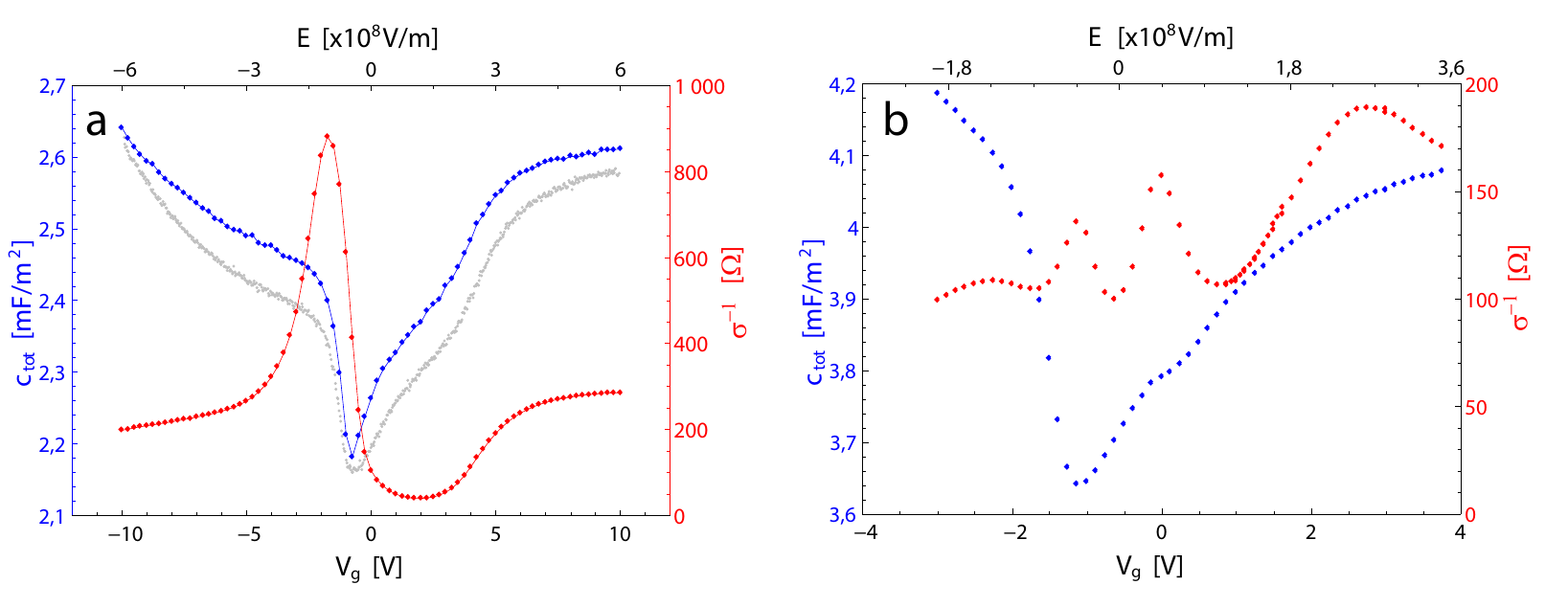}}
\caption{Additional evidence of Dirac screening breakdown at high electric field.
Panel a) Hysteresis loops of capacitance in sample $S_A$ for increasing gate voltage sweep
amplitudes $V_m$ (rainbow colors).  As seen in the inset, the width $\Delta V$ of
 hysteresis loops sets in at $V_g\geq3\;\mathrm{V}$.
 It corresponds to the resistance minimum in Fig.\ref{fig4}-a
identified as the occurrence of the first VP state substantially and denoted as VP$_1$ above.
Note also that the capacitance minima are shifted upward in the large  $V_g^m$ loops, indicating the presence of bulk carriers.
Panel b) longitudinal resistance $R_{xx}$  and Hall resistance $R_{xy}$ (inset) measured in a type-A  Hall bar.
A secondary resistance peak is observed in $R_{xx}$, similar to that in Fig.\ref{fig4}-a, which is accompanied by an S-shape
 anomaly in $R_{xy}(B)$ (inset). The anomaly is highlighted in the  $d R_{xy}/d B(V_g)$ waterfall plot in panel c),
 where data are fitted by an empirical function $\alpha/\cosh^2(B/B_0)$ (solid lines).  As the anomaly onset coincides
 with the resistance minimum (red line), we conclude that a second type of carrier does nucleate at the conductivity maximum denoted as VP$_1$.}\label{fig5}
\end{figure}

\emph{VPS, charge metastability and Dirac screening breakdown}.
In Fig.\ref{fig5} we analyze two complementary aspects of the
TI charging phenomenology. Fig.\ref{fig5}-a shows a set of
hysteretic charging characteristics $C_{tot}(V_g)$ for
increasing gate voltage amplitudes $V_g^m$ in sample $S_A$.
They have a butterfly shape with two capacitance minima that
are shifted upward or backward depending on the direction of
the voltage sweep. We attribute this metastability to the
breakdown of Dirac screening and the nucleation of bulk
carriers at high electric field, either electrons or holes
depending on the field history. The loop width, $\Delta
V_g(V_g^m)$ (inset) sets-in at $V_g^m\simeq3\;\mathrm{V}$ and
gradually increases up to $\Delta V_g\simeq3\;\mathrm{V}$ at
$V_g^m\simeq10\;\mathrm{V}$, corresponding to a bulk charge
variation $\Delta n_B=\pm 2\;10^{12}\;\mathrm{cm^{-2}}$. Note
that the capacitance minima are shifted upward at large $V_g^m$
supporting the contribution of additional bulk carriers in the
compressibility. The presence of charge metastability signals
the Dirac screening breakdown.

\begin{figure}[ht]
\centerline{\includegraphics[width=16cm]{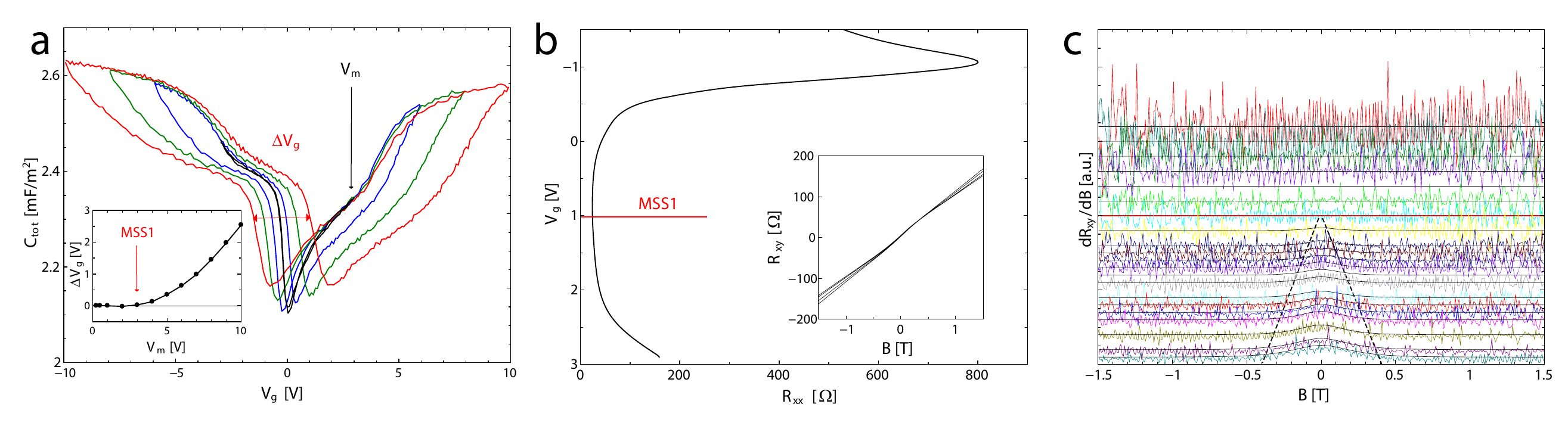}}
\caption{High electric field capacitance and resistivity measurements in samples  $S_A$ (panel a)
and  $S_B$ (panel b). The low-frequency data (gray dots in panel a) are added for comparison.
To overcome hysteresis, the data are plotted for an increasing gate voltage.
The most prominent features are the apparition of additional resistance peaks in the electron regime,
the Dirac peak being signaled by the capacitance dip. In sample $S_A$ the resistance peak is
accompanied by a bump in the capacitance.}\label{fig4}
\end{figure}

\emph{Hall signature of VPS carriers}. Figs.\ref{fig5}-b,c show
the $R_{xy}$, $R_{xx}$ magneto-transport measurements performed
in a type-A Hall bar. The capped sample shows a clear maximum
of the sheet resistance accompanied by a change in the sign of
the Hall resistance (not shown) indicating the state with
lowest carrier density at $V_g\approx -1.0\;\mathrm{V}$.
Increasing the gate voltage (electron side) decreases the sheet
resistivity $\sigma^{-1}$ up to $V_g\approx 1.0\;\mathrm{V}$.
For higher gate voltages $\sigma^{-1}$ increases monotonically
with $V_g$ (Fig.\ref{fig5}-c) and a non-linearity in the Hall
resistance develops for low magnetic fields (inset of
Fig.\ref{fig5}-b). The uncapped sample consistently shows
additional peaks in the longitudinal resistance (not shown),
with the only difference, that the p-conducting regime could
not be reached in the available gate voltage range due to the
high (unintentional) n-doping. Beside the Dirac peak (at
$V_g=-1.1\;\mathrm{V}$), $R_{xx}(V_g)$ exhibits a secondary
peak (at $V_g\simeq 3\;\mathrm{V}$) reminiscent of the one
observed in the capacitor measurement of Fig\ref{fig4}-a. The
Hall resistance $R_{xy}\propto B$ develops an anomaly  at
$V_g\gtrsim1\;\mathrm{V}$ (inset of Fig.\ref{fig5}-b) signaling
the advent of a secondary carrier type. The anomaly is best
depicted by plotting the derivative $\partial R_{xy}/\partial
B$ in Fig.\ref{fig5}-c, where the base line is shifted
according to the applied gate voltage. Data are fitted using an
empirical function $\alpha/\cosh^2(B/B_0)$ (solid lines) used
to extract the magnetic field range $B_0$ of the anomaly. We
find $B_0\propto(V_g-V_{g1})$ with $V_{g1}=1.0\;\mathrm{V}$
corresponding to the resistance minimum in that sample. This
shows that the scattering peak is indeed accompanied by the
nucleation of a second carrier type, consistently with the
VPS-subband interpretation. A similar analysis has been carried
out in type-B samples where we observe multiple $R_{xx}$ peaks
(not shown), also reminiscent of the capacitor measurement in
Fig.\ref{fig4}-b.

\subsection{The VPS phase diagram}

The heuristic model below  predicts a  VP state series,
$E_{m}=\sqrt{m}E_{1}$. In sample $S_A$ the second VPS, at
$E_{2}\simeq0.5\;\mathrm{eV}$, is at the limit of experimental
reach and deep in the metastable regime. This difficulty is
circumvented in sample $S_B$ which has a finite n-type chemical
doping and smaller $E_{m}$'s due to a smaller capping bandgap
$\Delta_2$ (see below). As shown in Fig.\ref{fig4}-b the
scattering spectroscopy is even richer: besides the Dirac peak
identified by the capacitance dip, we observe \emph{two
resistance peaks} in the electron side and \emph{one faint
resistance peak} in the hole side. Similarly to sample $S_A$
(Fig.\ref{fig4}-a), the conductance peaks (resistance dips) are
not accompanied by capacitance features.

For direct comparison with theory we provide in Figs.\ref{fig6}
a summary of our experimental observations. Figs.\ref{fig6}-a,c
show the quantum capacitance and conductivity as a function of
the TI charge density for both samples. We identify the Dirac
peak position by the coinciding capacitance/conductance minima,
and the VP states by the conductance maxima. These peaks are
reported in the density-electric field $n(\mathcal{E}_{ins})$
diagrams of Figs.\ref{fig6}-b,d which facilitates contact with
more standard measurements where the surface chemical potential
is not accessible experimentally. Note that $\mathcal{E}_{ins}$
is the surface electric field at the insulator side, which
differs from the HgTe-TI side by the ratio
$\varepsilon_{ins}/\varepsilon_{HgTe}$ of permittivities. The
green lines in Figs.\ref{fig6}-b,d are theoretical fits with
the model below. As mentioned before, an advantage of this
$n(\mathcal{E}_{ins})$ diagram lies in the fact that the
capacitor charging law reduces to a straight line. Actually,
this is strictly true for the total TI charge, including bulk
carriers. Later on when comparing with theory, one should keep
in mind that the theoretical phase diagram holds only for the
surface state density.

\section{\label{theory} Theory}

In this section, we describe the topological-normal junction
giving rise to surface states over a penetration depth $\xi$
within a simplified effective four-band model and a gradual
interface of length $\xi$ between the topological
(inverted-gap) and the trivial (normal-gap) insulator. Beside
the Dirac fermion we find a set of degenerate massive surface
states and we study the effect of an applied electric field on
the surface-states spectra. Our theoretical modeling is
 corroborated with numerical $k \cdot P$ calculations of a gradual junction using six-band
models particularly designed for HgTe.

\subsection{\label{effectivetheory} Effective model of surface states}
In order to model the insulating phases, we use the simplified
linear four-band $k\cdot P$-Hamiltonian describing the bands
around the $\Gamma$-point of an inverted band structure
[\onlinecite{qizhang2011review}]
\begin{equation}\label{eq:hamiltonian}
	\hat{H}_{0}({\bf k}, \Delta) = \left[
		\begin{array}{cccc}
			\Delta &\hbar v_F k_y & 0 & \hbar v_F(k_z-ik_x)\\
			\hbar v_F k_y & -\Delta & \hbar v_F(k_z-ik_x) & 0\\
			0 & \hbar v_F(k_z+ik_x) & \Delta & -\hbar v_F k_y\\
			\hbar v_F(k_z+ik_x) & 0 & -\hbar v_F k_y & -\Delta
		\end{array}
	\right],
\end{equation}
for which the spectrum consists of two doubly degenerate bands
$\varepsilon_{k}^{(\pm)} = \pm \sqrt{\Delta^2 + \hbar^2v_F^2
k^2}$. This spectrum is independent of the sign of the gap
parameter $\Delta$. In the following we model HgTe as an
inverted insulator of gap $ -\Delta_1 < 0$ that is in contact
with a normal insulator of gap $\Delta_2 > 0$. The
corresponding bulk Hamiltonians are $\hat{H}_{0}({\bf k},
-\Delta_1)$ and $\hat{H}_{0}({\bf k}, \Delta_2)$ as represented
on the left- and right-hand sides of panel (a) of Fig.
(\ref{fig7}).

We consider that HgTe is located in a region $z < 0$ and that
the normal insulator (CdHgTe or HfO$_2$) is located at $z > \xi
$. We model the interface $0 < z < \xi$ between the two
semi-conductors using an interpolating Hamiltonian
$\hat{H}_{s0} = \hat{H}_0\left[{\bf k}, -\Delta_1 +
e\mathcal{E}_T z \right]$ with the characteristic field
\begin{equation}
\mathcal{E}_T = \frac{\Delta_1+\Delta_2}{e\xi}\; . \label{ET}\end{equation} This field plays the
role of a confinement or gap field that we have chosen to have
the same physical dimension as an electric field. The evolution
of the gap is sketched as a green line in Fig.~\ref{fig7}-a and
can be viewed as a three-dimensional generalization of the
procedure described in Ref.[\onlinecite{Karzig2013prx}] for a
one-dimensional system and in Ref.
[\onlinecite{tchoumakovarxiv}] for Weyl semimetals. The
Hamiltonian contains a pair of non-commuting variables $[z,
k_z] \neq 0$ that can be merged into the same matrix elements
with the help of a $k-$independent unitary transformation
$|\Psi\rangle = \hat{U}|\Psi'\rangle$. One finds in this new
basis
\begin{equation}\label{eq:rotation0}
	\hat{H}_{s0}' = \hat{U}^{\dagger} \hat{H}_{s0} \hat{U}
	=
	\left[
		\begin{array}{cccc}
			\hbar v_F k_x & \hbar v_F k_y & 0 & \sqrt{2\hbar  v_F e\mathcal{E}_T}\hat{a}^{\dagger}\\
			\hbar v_F k_y & - \hbar v_F k_x & \sqrt{2\hbar  v_F e\mathcal{E}_T} \hat{a}^{\dagger} & 0\\
			0 & \sqrt{2 \hbar v_F e\mathcal{E}_T} \hat{a} & \hbar v_F k_x & - \hbar v_F k_y \\
			\sqrt{2 \hbar v_F e\mathcal{E}_T} \hat{a} & 0 & -\hbar v_F k_y & - \hbar v_F k_x
		\end{array}
	\right],
\end{equation}
where we have introduced the ladder operators $\hat{a} =
\left[\hbar v_F k_z - i \left( e\mathcal{E}_T z - \Delta_1
\right) \right]\left/\sqrt{2 \hbar v_F e\mathcal{E}_T }\right.$
and $\hat{a}^{\dagger} = \left[ \hbar v_F k_z + i \left(
e\mathcal{E}_T z - \Delta_1 \right) \right]\left/\sqrt{2\hbar
v_F e\mathcal{E}_T }\right.$ such that $\left[ \hat{a},
\hat{a}^{\dagger} \right] = 1$. This Hamiltonian is similar to
that in a magnetic field [\onlinecite{markreview}] and we
introduce the number states $|m\rangle$ associated to the
number operator $\hat{m} = \hat{a}^{\dagger}\hat{a}$, with
$\hat{m}|m\rangle = m |m\rangle$. These states are localized at
the interface between two insulator with a mean position
$\langle z \rangle = \Delta_1/e \mathcal{E}_T = \Delta_1
\xi/(\Delta_1+\Delta_2) \in [0,\xi]$.

For $m \geq 1$ one finds an infinite number of states of the
form $|\Psi_{m,\bf{k}}\rangle = \left(
			\alpha_1 |m\rangle,
			\alpha_2 |m\rangle,
			\alpha_3 |m-1\rangle,
			\alpha_4 |m-1\rangle
	\right). $ Their band dispersion is doubled in that one
finds, for each value of $m$, a band at positive and negative
energy,
\begin{equation}
 \varepsilon_{m,k_x,k_y}^{(\pm)} = \pm \sqrt{\hbar^2v_F^2(k_x^2 +
k_y^2) + 2 \hbar v_F e\mathcal{E}_T m},
\end{equation}
which yields the relevant ${\bf k}=0$ separation
\begin{equation}\label{eq:VP}
 E_{\pm m}=\varepsilon_{m,k_x=k_y=0}^{(\pm)} = \pm \sqrt{2 \hbar v_F e\mathcal{E}_T m}=
 \pm \sqrt{2 \hbar v_F (\Delta_1+\Delta_2) m/\xi}
\end{equation}
between the VPS, as already mentioned in Sec.\ref{experiment}.B
in the estimation of the surface width $\xi$. In addition,
these states, which were first identified by Volkov and
Pankratov in 1985 [\onlinecite{Volkov1985jetp}], are doubly
degenerate and depend explicitly on the characteristic field
(\ref{ET}) and thus on the parameters characterizing the
interface. In the limit of a sharp surface, i.e. $\xi
\rightarrow 0$ so that $\mathcal{E}_T \rightarrow \infty$,
these states are shifted to high energies and do not play any
physical role.

In contrast to theses bands, the $m = 0$ surface states are not
degenerate. They are of the form $|\Psi_{0, {\bf k}}\rangle =
\left( \alpha_1 |0\rangle, \alpha_2 |0 \rangle, 0, 0 \right)$
and their energy dispersion is that of a two-dimensional Dirac
cone $\varepsilon_{0,k_x,k_y}^{(\pm)} = \pm \hbar v_F
\sqrt{k_x^2 + k_y^2}$. It is independent of $\mathcal{E}_T$ and
one can show [\onlinecite{tchoumakovarxiv}] that this state is
of topological nature. This topological surface state survives
in the limit of an infinitely sharp interface, as expected, and
shows a dispersion that only depends on the bulk parameter
$v_F$, in agreement with previous studies
[\onlinecite{qizhang2011review}]. We represent the spectra of
the VPS$_{\pm m}$ states and the Dirac state in panel (a) of
Fig.~\ref{fig7}.

Most saliently, the VPS can be modified to great extent by an
electric field applied perpendicular to the surface. As an
effect of charge screening, the associated electrostatic
potential drops in the interface and generates and electric
field  $\boldsymbol{\mathcal{E}} \equiv -\mathcal{E} {\bf e}_z
\approx -V_0/\xi {\bf e}_z$, where $V_0=\mu/e$ is the surface
potential. The interface Hamiltonian, for $0 < z < \xi$, is now
$\hat{H}_s = \hat{H}_{s0}' + (V_0 - e \mathcal{E} z)
\mathbbm{1}$. Notice that the added term remains invariant
under the above-mentioned rotation that leads to the form of
the Hamiltonian given by Eq.~(\ref{eq:rotation0}). As detailed
in App.\ref{sec:lorentz}, the spectrum now depends on the ratio
$\beta = \mathcal{E}/\mathcal{E}_T$ between the applied and the
characteristic electric fields. The spectrum for the $m \geq 1$
states is still doubly degenerate with the dispersion relations
$\varepsilon_{m,k_x,k_y}(\beta) =
\sqrt{(1-\beta^2)\hbar^2v_F^2(k_x^2 + k_y^2) + 2m
(1-\beta^2)^{3/2} \hbar v_F e\mathcal{E}_T}$. One observes that
the gap (\ref{eq:VP}) is reduced by the applied electric field
according to
\begin{equation}
E_{\pm m}(\mathcal{E}) = \pm \sqrt{m}\;\sqrt{2\hbar v_F e\mathcal{E}_T}\;(1-\mathcal{E}^2/\mathcal{E}_T^2)^{3/4} \;,
\label{excitation-energy}\end{equation}   and that the critical surface density obeys
\begin{equation}
n_m(\mathcal{E}) = \frac{m(m+1)}{2}\;\frac{e\mathcal{E}_T}{2\pi \hbar v_F}\;(1-\mathcal{E}^2/\mathcal{E}_T^2)^{1/2}\;. \label{excitation-density}\end{equation} At the same time
the dispersion relation flattens out because of the reduced
Fermi velocity that vanishes at $\mathcal{E}_T$ according to
\begin{equation}
v_F(\mathcal{E}) = v_F \;(1-\mathcal{E}^2/\mathcal{E}_T^2)^{1/2}\;. \label{softening}\end{equation}
The special $m = 0$ surface state is also flattened and one
finds $\varepsilon_{0,k_x,k_y}^{(\pm)} = \pm (1-\beta^2)^{1/2}
\hbar v_F \sqrt{k_x^2 + k_y^2}$.

In panel (b) of Fig.~\ref{fig7} we represent the extrema of the
VPS$_{\pm m}$ (in green) and of the Dirac state at $k_x=k_y=0$
(in red) as a function of $\beta = \mathcal{E}/\mathcal{E}_T$.
We observe that
 VPS merge for an electric field close to the critical field $\mathcal{E}_T$. Beyond this limit
our model (\ref{eq:Hsfinal}) has no bound state so that the
interface behaves as a conventional semiconducting
heterojunction. This shows that $\mathcal{E}_T$ is not only a
characteristic field governing the massive surface state
spectrum but actually a genuine critical field for the
topological nature of the interface. The applied electric
potential also influences the chemical potential [blue line in
Fig.~(\ref{fig7})] that will eventually cross the VPS. This
leads to the experimentally observed kinks in the
compressibility and features in the conductivity that we
discuss in detail in the following section.

We finish this theoretical section with a discussion of the
density of states $\rho(\epsilon)$, which is directly
proportional to the quantum capacitance (at $T=0$). The density
of states associated with the surface states per unit area
reads
\begin{equation}\label{eq:dos}
	\rho(\varepsilon) = \frac{|\varepsilon|}{2\pi \hbar^2 v_F^2(1-\beta^2)}   \sum_{l} \Theta\left[|\varepsilon| - (1-\beta^2)^{3/4} \sqrt{2 \hbar v_F e\mathcal{E}_T m}\right],
\end{equation}
where $\Theta(x)$ is the Heaviside function. The corresponding
behavior of the quantum capacitance is represented in panel (c)
of Fig.~\ref{fig7} for typical values of $\beta =
\mathcal{E}/\mathcal{E}_T$. One observes that the gaps are
smaller and that the density of states becomes enhanced for
larger electric fields (in red). This is a direct consequence
of the reduced Fermi velocity in the presence of an electric
field, as pointed out above. Notice, however, that in the
experimental setup the chemical potential depends itself on the
applied electric potential and therefore jumps from one curve
to the other. Moreover, we expect  this density of state to be
smeared in the presence of disorder.

\subsection{\label{numericaltheory} Numerical $k \cdot P$ treatment of surface states}

To complement the previous analytical study, we have performed
a numerical study of the band structure of an HgTe/CdTe
interface, based on a $k.P$ model. Our numerical approach
amounts to discretizing a standard Kane model for the $6$ bands
$\Gamma_{6,\pm 1/2} , \Gamma_{8,\pm 1/2},\Gamma_{8,\pm 3/2}$
hamiltonian with parameters for HgTe and CdTe known from the
literature [\onlinecite{Novik2005prb}]. The parameters of the
model is interpolate between their values in both materials
over a distance $\xi$, corresponding to the size  of the
interface. This description incorporates the stress induced by
the lattice mismatch through a Bir-Pikus term. The resulting
band structure is shown in Fig.\ref{fig8}, where the color
encodes the eigenstate's density around the HgTe/CdTe
interface. The band structure is calculated for an HgTe
thickness of $70\;\mathrm{nm}$ and $\xi=5\;\mathrm{nm}$. Note
that this band structure is obtained at zero electric field and
using CdTe ($2\Delta_2\simeq1.5\;\mathrm{eV}$) as a capping
layer boundary. We find a massive surface subband at
$1\;\mathrm{eV}$  which is accompanied by a strong depletion of
the bulk state amplitude in the surface layer that confirms the
above picture of high energy surface states (VPS). The
excitation energy is quite large (still smaller than
$2\Delta_2$) and found to be sensitive to the detail of the
shape of the smooth interface. This is consistent with the
analytical approach of section \ref{effectivetheory} which
focussed on a linear interpolation between the two materials
and neglects asymmetry between the two materials beyond the gap
inversion. We stress that these VPS are predicted at zero
electric field, and assume that they follow the electric field
dependence predicted by the approach of section
\ref{effectivetheory}, as corroborated by Ref.
[\onlinecite{Tchoumakov2017arXiv}].

 \section{Comparison with experiment}\label{comparison}

\begin{figure}[ht]
\centerline{\includegraphics[width=12cm]{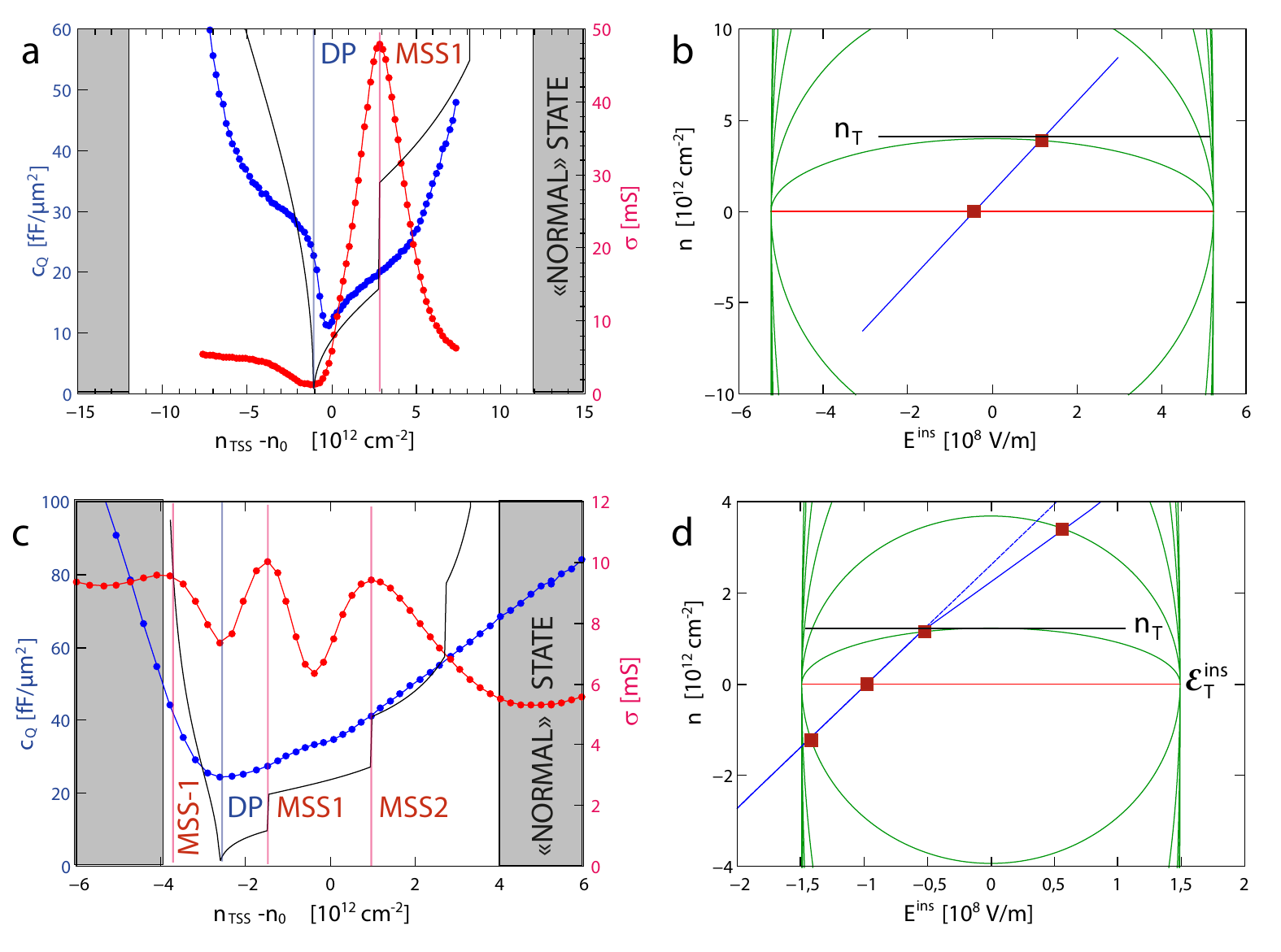}}
\caption{Density - electric field phase diagram of strained bulk HgTe.
The compressibility-conductance data (blue-red data points) from samples $S_A$ (panel a)
and $S_B$ (panel c), are compared to the prediction of the model in Section \ref{theory}.
By adjusting the resistance peak positions with respect to the Dirac point we deduced the model
parameters (see the text). On panel b) and d) the green solid lines correspond to the theoretical
predictions of energies of the surface states. Square signals correspond to energies of VP states
extracted from experiment. The MITI-Cap charging paths are included in the $n$--$\mathcal{E}$
phase diagram as depicted in panels b) and d). These phase diagrams are obtained from the
$\mathcal{E}$-behavior of the VP surface states (\ref{excitation-energy}) plotted in Fig. \ref{fig7}-b (see
theory section \ref{theory}). From this analysis we deduce the critical
electric field of the THJ beyond which surface states disappear alltogether: $\mathcal{E}_{T}^A=2.64\;10^8\;\mathrm{Vm^{-1}}$ and
$\mathcal{E}_{T}^B=0.81\;10^8\;\mathrm{Vm^{-1}}$. The smaller $\mathcal{E}_{T}$
in $S_B$ is due to the absence of a capping layer. Above $\mathcal{E}_{T}$,
the topological-trivial insulator interface behaves as a normal, non-topological, interface.
Two such subbands can be seen in $S_B$ due
to chemical doping and a lower excitation gap.}\label{fig6}
\end{figure}

\begin{figure}[ht]
\centerline{\includegraphics[width=12cm]{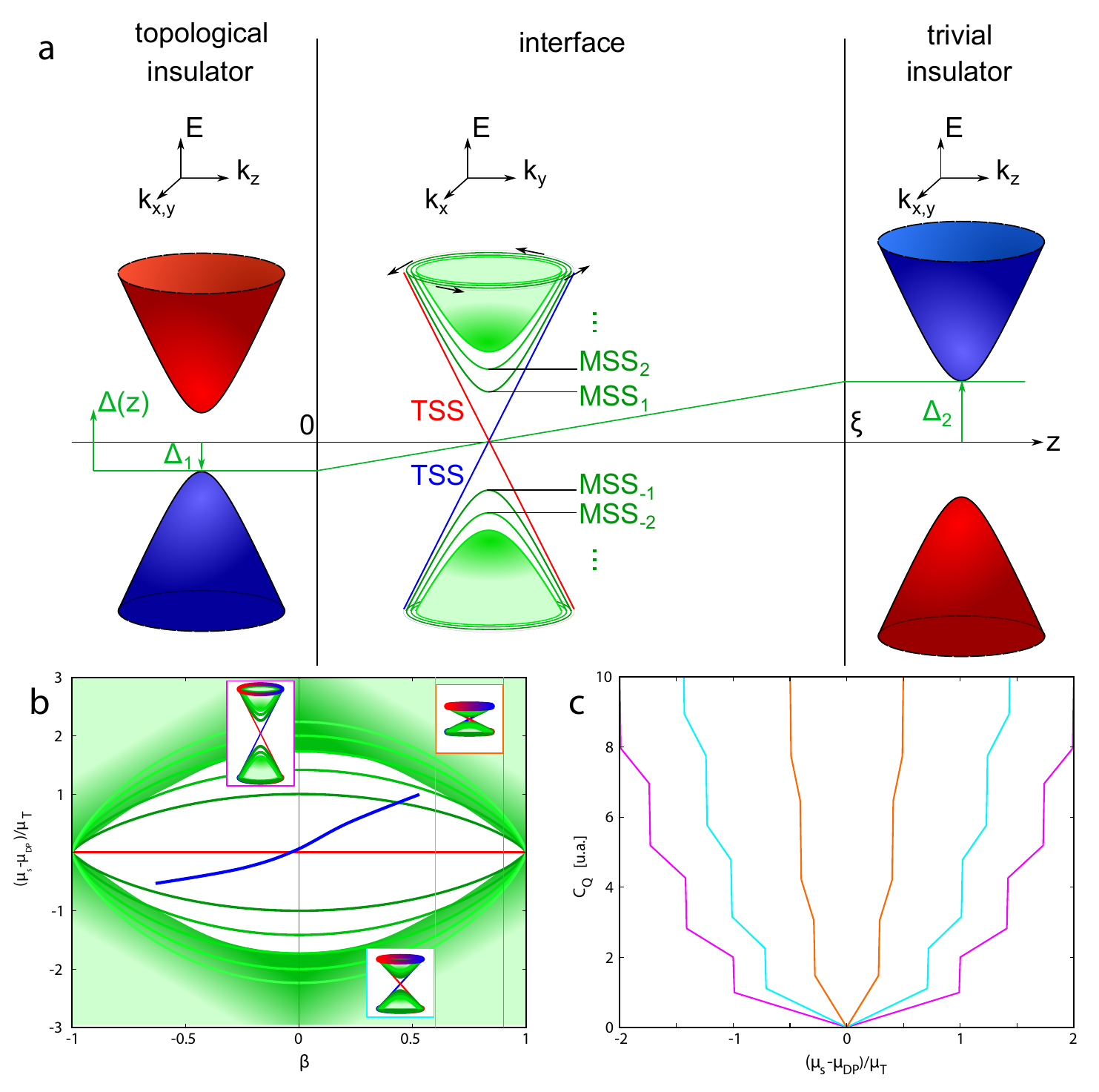}}
\caption{Model of massive surface states (VPs) in topological insulators.
Panel a) Simplified model for the interface of a topological insulator (on the left)
and a normal insulator (on the right). In the interface one observes multiple surface states.
Panel b) Band gaps of the Dirac and VP states as a function of the reduced electric field
$\beta = \mathcal{E}/\mathcal{E}_T$. The blue solid line is a sketch of a capacitor charging
line measured in Fig.\ref{fig3}-c. Panel b) is equivalent to the $n-\mathcal{E}$ phase diagram of
Fig. \ref{fig6}-b and -c.
Panel c) Illustration of the double effect of electric
field in the quantum capacitance $C_Q(\mu,\mathcal{E})$ for selected values of the parameters
represented by colored lines in panel b). The group velocity of the Dirac fermion decreases with
increasing electric field and its density of states rises up to the critical field with a
vanishing of the Fermi velocity $C_Q(\mu)_{\mathcal{E}\rightarrow\infty}$  at the critical
field $\mathcal{E}_T$. At finite doping the subbands cross the Fermi energy at
$\mathcal{E}<\mathcal{E}_T$ giving rise to a stepwise increase of $C_Q(\mathcal{E})_{\mu=Cte}$. }\label{fig7}
\end{figure}

\begin{figure}[ht]
\centerline{\includegraphics[width=7cm]{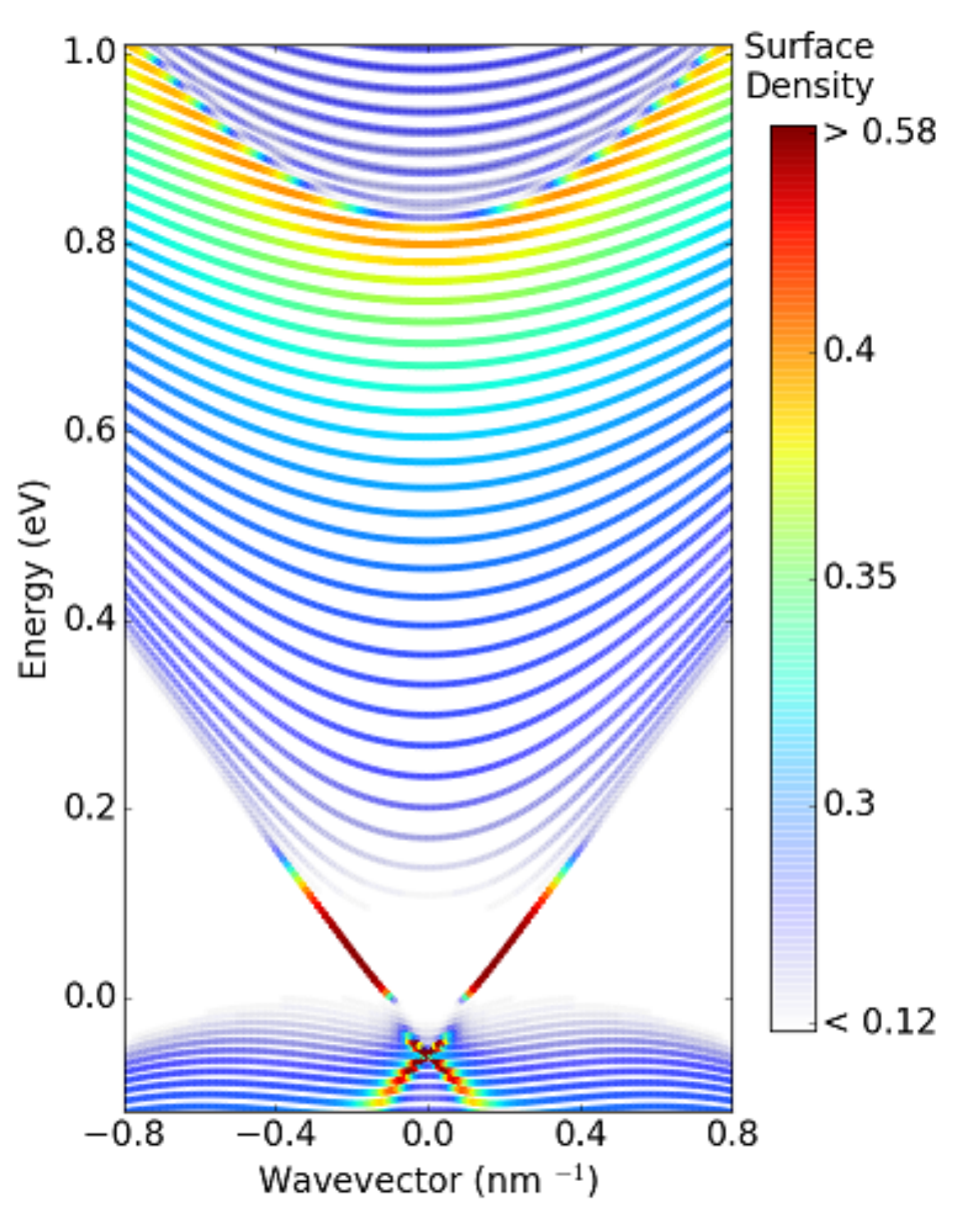}}
\caption{Numerical dispersion relation $E(k_\parallel)$ for an HgTe/CdTe interface obtained
within a 6-band Kane model. $k_\parallel$ corresponds to the momentum along the interface.
The color encodes the density of eigenstates in  a region of $\sim 6\;\mathrm{nm}$ around the
interface. The existence of localized states around the interface is shown within the gap but
also at high energy around $1\;\mathrm{eV}$.    }\label{fig8}
\end{figure}

The main evidence of VPS is the scattering peak observed in the
intrinsic sample $S_A$ (Fig\ref{fig2}-e and Fig\ref{fig6}-a),
where the absence of classical massive surface state and the
Dirac screening of bulk states have been previously warranted
experimentally [\onlinecite{Brune2014prx}]. The measured VPS1
gap $E_1=0.35\;\mathrm{eV}$ is close to the theoretical value
$E_{1}(0)\simeq0.4\;\mathrm{eV}$, deduced from
Eq.(\ref{excitation-energy}), and is accompanied by a second
carrier type observed in the Hall bar measurements. In addition
we report that the VPS1 triggers the onset of charge
metastability signaling the breakdown of Dirac screening
[\onlinecite{Brune2014prx}].

The second evidence is the observation of a series of
scattering peaks in the uncapped sample $S_B$ (Fig\ref{fig4}-b
and Fig\ref{fig6}-c). The capping of sample $S_B$ being
ill-defined, we cannot make a direct comparison of VPS energy
with theory so that $\Delta_2$ becomes a fitting parameter.
Still we can adjust the peaks series with the predicted
sequence $n_{m} \propto m(m+1)/2$ of
Eq.(\ref{excitation-density}).

The third evidence relates to the electric field red-shift. As
seen in Fig\ref{fig6}-d, the agreement of VPS spectroscopy with
theory involves a $\sim10\%$ electric field red-shift for the
VPS2 state of sample $S_B$ consistent with
Eq.(\ref{excitation-energy}). As for the VPS-1 state, its
observation fully relies on a strong field renormalization of
the gaps, meaning that VPS-1 peak position signals the vicinity
of the critical field $\mathcal{E}_T$. The fact that VPS-1 is
smeared is also consistent with theory since the Fermi velocity
vanishes at $\mathcal{E}_T$ according to Eq.(\ref{softening}).
Moreover the conductivity becomes featureless above
$\mathcal{E}_T$ also in agreement with theory. Our scattering
spectroscopy measurements thus support the field-effect induced
red-shift and suppression of VPS  predicted in Section
\ref{theory}. Although a deeper insight into the field
suppression is desirable, it remains very challenging as a full
mapping of the $n(\mathcal{E})$ diagram would require an
in-situ tuning of the insulator permittivity and/or TI chemical
doping. An extension of this work can be envisioned using a
series of similar HgTe THJs with varied capping bandgap
$\Delta_2$ and/or insulator permittivity $\varepsilon_{ins}$.

Before concluding, let us detail our analysis of the
$n(\mathcal{E})$ phase diagrams in Figs.\ref{fig6}-b,d.
Eq.(\ref{excitation-density}) predicts sub-band minima for a
surface state density $n_m=n_1\times
\frac{m(m+1)}{2}\sqrt{1-\beta^2}$ where
$n_1=e\mathcal{E}_T/2\pi \hbar v_F$ and
$\beta=\mathcal{E}_{HgTe}/\mathcal{E}_T=\varepsilon_{ins}\mathcal{E}_{ins}/\varepsilon_{HgTe}
\mathcal{E}_T$. Taking
$\mathcal{E}_T=1.2\;10^8\;\mathrm{V/m}$,
$\varepsilon_{ins}^A=C_{ins}t_{ins}/\varepsilon_{0}\simeq4.5$
and $\varepsilon_{HgTe}\simeq20$
[\onlinecite{Baars1972ssc,Capper2010}], we estimate for sample
$S_A$ the applied critical field
$\mathcal{E}_c^A=\varepsilon_{HgTe}\mathcal{E}_T/\varepsilon_{ins}^A\simeq
5\;10^{8}\;\mathrm{V m^{-1}}$  and a surface critical density
$n_1^A=2.9\;10^{12}\;\mathrm{cm^{-2}}$ (with
$v_F=10^6\;\mathrm{m/s}$). Direct total charge measurement
gives a larger $n_1\simeq4\;10^{12}\;\mathrm{cm^{-2}}$ due to
the presence of bulk carriers. The same analysis can be carried
out for the doped sample $S_B$ where we obtain a quantitative
agreement for the position of three VPS peaks (VPS-1, VPS2 and
VPS3) with a single parameter
$\mathcal{E}^{B}_T=(0.4\pm0.1)\;10^{8}\;\mathrm{V m^{-1}}$,
yielding $n_1^B=1.\;10^{12}\;\mathrm{cm^{-2}}$. The fact that
$\mathcal{E}^{B}_T\simeq\mathcal{E}^{A}_T/3$ highlights the
importance of a capping layer in strengthening the robustness
of surface topological states.

Finally we have plotted in Figs.\ref{fig6}-(a,c) (solid black
lines) the theoretical prediction for the quantum capacitance
calculated from the density of states of surface states in
Eq.(\ref{eq:dos}). While we see the linear increase of a
secondary carrier density in the Hall bar measurements, we  do
not resolve the compressibility steps signaling the onset of
massive subbands. A possible reason for this discrepancy is the
presence of residual bulk carriers, obscuring the
compressibility resolution and explaining the compressibility
background observed in Fig.\ref{fig3}-d. These carriers can be
disorder-induced small scale charge puddles or more likely, in
our high mobility samples, large-scale puddles induced by the
non-uniformity of the applied electrostatic surface potential.
Considering the small transport bandgap of HgTe, the potential
uniformity requirements are especially stringent.

\section{Conclusion}\label{conclusion}

Our comprehensive set of measurement, complemented by a
heuristic model, support the existence of intrinsic massive
surface states,  the Volkov-Pankratov states, accompanying
Dirac states at the interface between TI HgTe and the insulator
CdHgTe.  Such high-energy topological states, that are
intrinsic to topological matter, have long been predicted
[\onlinecite{Volkov1985jetp,Karzig2013prx}] but remained
elusive until now. Our work is the first report and systematic
study of these states, and we hope it will trigger further
investigations into different materials such as the Bi-based TI
[\onlinecite{Inhofer2017arXiv}] or different topological phases
such as massive Majorana states, helical edge states or Fermi
arcs. As pointed out in Ref.[\onlinecite{Karzig2013prx}] and
demonstrated in this experiment, massive surface states play an
important role in restricting the phase diagram where
topological protection is robust such as the Dirac screening
[\onlinecite{Karzig2013prx}]. This phase diagram can be
enriched by applying an external magnetic field, or simply by
increasing the temperature, highlighting the possible role of
electron-electron interactions.

\section{\label{sec:lorentz} Appendix : Lorentz boost of surface states in an electric field }
In section \ref{effectivetheory} we have introduced the
Hamiltonian  $\hat{H}_s = \hat{H}_{s0}' + (V_0 - e \mathcal{E}
z) \mathbbm{1}$. In order to deal with this new $z-$dependence,
we perform a Lorentz boost on the time-independent
Schr\"odinger equation $e^{\frac{\eta}{2} \hat{J}
}(\hat{H}_{s}' - \varepsilon \mathbbm{1})e^{\frac{\eta}{2}
\hat{J} }|\bar{\Psi}\rangle = 0$, where the Lorentz boost
[\onlinecite{lukose2007prl}] is realized by the hyperbolic
transformation $|\Psi'\rangle = \mathcal{N} e^{-\frac{\eta}{2}
\hat{J} }|\Psi\rangle$ generated by $\hat{J} = \sigma_y \otimes
\sigma_x$, in terms of the Pauli matrices $\sigma_x$ and
$\sigma_y$.

In the case $\tanh(\eta) \equiv \beta =
\mathcal{E}/\mathcal{E}_T \in [-1,1]$ and $V_0 = -\beta
\Delta_1$, one can thus make the $(V_0- e \mathcal{E}
z)\mathbbm{1}$-term vanish in the co-moving frame of reference.
If we define the pseudo-magnetic field $B = \mathcal{E}_T/v_F$,
this transformation can be understood from the viewpoint of
\emph{special relativity} as a boost to a frame of reference
where the drift velocity $v_D = \mathcal{E}/B$ vanishes
[\onlinecite{lukose2007prl}]. The condition
$\mathcal{E}/\mathcal{E}_T = v_D/v_F \in [-1,1]$ is similar to
the existence of a limiting velocity, the Fermi velocity $v_F$
plays the role of the speed of light in special relativity. The
Schr\"odinger equation becomes $\hat{H}_{s}'|\bar{\Psi}\rangle
= \varepsilon |\bar{\Psi}\rangle$ with
\begin{equation}\label{eq:Hsfinal}
	\hat{H}_s' =
	\left[
		\begin{array}{cccc}
			\hbar v'_F k_x & \hbar v'_F k_y & 0 & \sqrt{2 \hbar v'_F e\mathcal{E}'_T} \hat{b} \\
			\hbar v'_F k_y & -\hbar v'_F k_x & \sqrt{2\hbar  v'_F e\mathcal{E}'_T} \hat{b} & 0\\
			0 & \sqrt{2 \hbar v'_F e\mathcal{E}'_T} \hat{b}^{\dagger} & \hbar v'_F k_x & -\hbar v'_F k_y\\
			\sqrt{2 \hbar v'_F e\mathcal{E}'_T} \hat{b}^{\dagger} & 0 & -\hbar v'_F k_y & -\hbar v'_F k_x
		\end{array}
	\right],
\end{equation}
with $v_F' = \sqrt{1-\beta^2}v_F$, $\mathcal{E}_T' =
(1-\beta^2)\mathcal{E}_T$ and the ladder operators $\hat{b},
\hat{b}^{\dagger}$ where
\begin{align}
	&\hat{b} = \frac{1}{(1-\beta^2)^{1/4}\sqrt{2 \hbar v_F' e\mathcal{E}_T'}} \left\{ \hbar  v_F k_x - i \sqrt{1-\beta^2} \left[ e \mathcal{E}_T z - \left( \Delta_1 + \frac{\beta }{{\beta^2-1}}\varepsilon \right) \right] \right\},\\
	&\hat{b}^{\dagger} = \frac{1}{(1-\beta^2)^{1/4}\sqrt{2 \hbar v_F' e\mathcal{E}_T'}} \left\{ \hbar  v_F k_x + i \sqrt{1-\beta^2} \left[ e \mathcal{E}_T z - \left( \Delta_1 + \frac{\beta }{{\beta^2-1}}\varepsilon \right) \right] \right\}.
\end{align}
As a consequence of the Lorentz boost, the ladder operators are
now explicitly energy-dependent as well as the mean position of
the number states $\langle z \rangle = \left.\left( \Delta_1 +
\frac{\beta }{{\beta^2-1}}\varepsilon \right)\right/
e\mathcal{E}_T$. For our theory to describe a surface state,
this position must be within $\langle z \rangle \in [0, \xi]$,
and one notices that for $\beta \rightarrow 1$ this condition
is not fulfilled for $\varepsilon \neq 0$.

The spectrum is found using states of the form $|\Psi_{m, {\bf
k}}\rangle = \left( \alpha_1 |m\rangle, \alpha_2 |l \rangle,
\alpha_3 |m-1 \rangle, \alpha_4 |m-1 \rangle \right)$ for $m
\geq 1$ and $|\Psi_{0, {\bf k}}\rangle = \left( \alpha_1
|0\rangle, \alpha_2 |0 \rangle, 0, 0 \right)$ for $m = 0$, in
the comoving frame of reference. The $|m\rangle$ states are the
eigenstates of the number operator $\hat{m} =
\hat{b}^{\dagger}\hat{b}$.

\begin{acknowledgments}

We gratefully acknowledge B.A. Bernevig, N. Regnault, R.
Ferreira, Y. Guldner, G. Bastard and M. Civelli for fruitful
discussions.

\end{acknowledgments}

\end{document}